\documentstyle[12pt]{article}
\begin{document}


\let\und=\b			
\let\ced=\c			
\let\du=\d			
\let\um=\H			
\let\sll=\l			
\let\Sll=\L			
\let\slo=\o			
\let\Slo=\O			
\let\tie=\t			
\let\br=\u			


\def\a{\alpha}
\def\b{\beta}
\def\c{\chi}
\def\d{\delta}
\def\e{\epsilon}		
\def\f{\phi}			
\def\g{\gamma}
\def\h{\eta}
\def\i{\iota}
\def\j{\psi}
\def\k{\kappa}			
\def\l{\lambda}
\def\m{\mu}
\def\n{\nu}
\def\o{\omega}
\def\p{\pi}			
\def\q{\theta}			
\def\r{\rho}			
\def\s{\sigma}			
\def\t{\tau}
\def\u{\upsilon}
\def\x{\xi}
\def\z{\zeta}
\def\D{\Delta}
\def\F{\Phi}
\def\J{\Psi}
\def\L{\Lambda}
\def\O{\Omega}
\def\P{\Pi}
\def\Q{\Theta}

\def\sfA{{\sf A}}                       \def\sfa{{\sf a}}
\def\sfB{{\sf B}}			\def\sfb{{\sf b}}
\def\sfC{{\sf C}}			\def\sfc{{\sf c}}
\def\sfD{{\sf D}}			\def\sfd{{\sf e}}
\def\sfE{{\sf E}}			\def\sfe{{\sf f}}
\def\sfF{{\sf F}}			\def\sff{{\sf f}}
\def\sfG{{\sf G}}			\def\sfg{{\sf g}}
\def\sfH{{\sf H}}			\def\sfh{{\sf h}}
\def\sfI{{\sf I}}			\def\sfi{{\sf i}}
\def\sfJ{{\sf J}}			\def\sfj{{\sf j}}
\def\sfK{{\sf K}}			\def\sfk{{\sf k}}
\def\sfL{{\sf L}}			\def\sfl{{\sf l}}
\def\sfM{{\sf M}}			\def\sfm{{\sf m}}
\def\sfN{{\sf N}}			\def\sfn{{\sf n}}
\def\sfO{{\sf O}}			\def\sfo{{\sf o}}
\def\sfP{{\sf P}}			\def\sfp{{\sf p}}
\def\sfQ{{\sf Q}}			\def\sfq{{\sf q}}
\def\sfR{{\sf R}}			\def\sfr{{\sf r}}
\def\sfS{{\sf S}}			\def\sfs{{\sf s}}
\def\sfT{{\sf T}}			\def\sft{{\sf t}}
\def\sfU{{\sf U}}			\def\sfu{{\sf u}}
\def\sfV{{\sf V}}			\def\sfv{{\sf v}}
\def\sfW{{\sf W}}			\def\sfw{{\sf w}}
\def\sfX{{\sf X}}			\def\sfx{{\sf x}}
\def\sfY{{\sf Y}}			\def\sfy{{\sf y}}
\def\sfZ{{\sf Z}}			\def\sfz{{\sf z}}


\def\ca{{\cal A}}
\def\cb{{\cal B}}
\def\cc{{\cal C}}
\def\cd{{\cal D}}
\def\ce{{\cal E}}
\def\cf{{\cal F}}
\def\cg{{\cal G}}
\def\ch{{\cal H}}
\def\ci{{\cal I}}
\def\cj{{\cal J}}
\def\ck{{\cal K}}
\def\cl{{\cal L}}
\def\cm{{\cal M}}
\def\cn{{\cal N}}
\def\co{{\cal O}}
\def\cp{{\cal P}}
\def\cq{{\cal Q}}
\def\car{{\cal R}}
\def\cs{{\cal S}}
\def\cat{{\cal T}}
\def\cu{{\cal U}}
\def\cv{{\cal V}}
\def\cw{{\cal W}}
\def\cx{{\cal X}}
\def\cy{{\cal Y}}
\def\cz{{\cal Z}}

\def\bd{\begin{displaymath}}
\def\ed{\end{displaymath}}
\def\be{\begin{equation}}
\def\ee{\end{equation}}
\def\ba{\begin{eqnarray}}
\def\ea{\end{eqnarray}}
\def\bea{\begin{array}}
\def\eea{\end{array}}
\def\bqn{\begin{eqnarray*}}
\def\eqn{\end{eqnarray*}}
\def\half{\frac{1}{2}}
\def\st{\star}
\def\v{\varphi}
\def\lbl{\label}
\def\dd{\partial}
\def\pa{\partial}
\def\rar{\rightarrow}
\def\itmb{\item[$\bullet$]}
\def\half{\frac12}
\def\intin{\int_{-\infty}^{+\infty}}
\def\pt{\phi^{(3)}}
\def\po{\phi^{(1)}}
\def\pf{\phi^{(5)}}

\begin{titlepage}
\title{
\vspace{-1cm}
\begin{flushright}
{\normalsize IFT/2/96}\\
{\normalsize hep-th/9601062}     
\end{flushright}
\vspace{1cm}
A Class of Exact Solutions of\\ the Wheeler -- De Witt Equation}
\author{
Jerzy Kowalski-Glikman\thanks{e-mail address:
jurekk@ift.uni.wroc.pl and jurekk@fuw.edu.pl}\\
Institute for Theoretical Physics\\
University of Wroc\sll{}aw\\
Pl.\ Maxa Borna 9, PL--50-205 Wroc\sll{}aw, Poland\\
and\\
Krzysztof A. Meissner\thanks{Supported in part by the Polish KBN 
Grant;\ 
e-mail address: meissner@fuw.edu.pl}\\
Institute of Theoretical Physics\\
Warsaw University\\
ul.\ Ho\.za 69, PL--00-681 Warsaw, Poland}
\date{January 1996}
\maketitle
\begin{abstract}
After carefully regularizing the Wheeler -- De Witt operator, which is
the Hamiltonian operator of canonical quantum gravity, we find a class
of exact solutions of the Wheeler -- De Witt equation. 
\end{abstract}
\end{titlepage}

The enigma of quantum gravity is probably the most challenging problem
of modern theoretical physics (for the recent reviews see \cite{Isham})
It is for the difficulty
of the problem
that in spite of its importance, very little progress has been made
so far. At the moment there are two major approaches to quantum gravity.
On the one hand, it is being
argued that gravity is a low energy
approximation of some drastically different fundamental theory;
the major examples of this kind of ideas are supergravities and,
more recently, superstring theories. If one accepts such a proposal,
"quantization" of classical general relativity does not
make much sense.

There is however another approach in attempt to quantize
the classical Einstein general theory of relativity. Nowadays the main
line of attack lies in using canonical quantization technique. In recent 
years a great
deal of excitement was risen by introduction of Ashtekar variables which
were expected to simplify the technical problems of older
canonical techniques. However, it seems that these hopes have not been 
fulfilled. Even though the use of the loop variables solves part of
constraints automatically, it turned out that the remaining ones are as
difficult to solve as the constraints of the traditional canonical
approach, based on the Wheeler -- De Witt equation (given
the result of the present paper rather more difficult.) Moreover,
due to the fact that existing solutions are expressed in terms of very
exotic variables, their physical meaning is rather obscure. This is the
reason why we choose to analyse again the original approach to quantum
gravity which dates back to seminal papers of Wheeler and De Witt
\cite{Wheeler}, \cite{DeWitt}. It should be stressed however, that it
seems likely that all these three approaches may reveal three different
regimes of the same ultimate theory (see \cite{Isham} for extensive
discussion of this point).

The Wheeler -- De Witt approach to quantization of general relativity is
based on the following set of equations describing ``the wavefunction of
the universe":
\be
\nabla_m {\sf p}^{mn}(x)\, \Psi(h_{mn}) = 0 
\label{diff}
\ee
\be
{\sf H}_{WDW}\Psi = \left(-{1 \over \mu} \frac{{\cal G}_{klmn}}{ 
\sqrt h} {\sf p}^{kl} \sfp^{mn} + \mu \sqrt h \left(2 \Lambda +  R 
\right)\right)\Psi =0 , \label{ham}
\ee
where
$\mu = \left( 16 \pi G \right)^{- 1}$, $G$ is the Newton's constant,
\be
{\cal G}_{klmn} ={\frac 12}
\left(h_{km} h_{ln} + h_{kn} h_{lm} - h_{kl} h_{mn} \right), \label{wdwmet}
\ee
${\sf p}^{kl}$ are momentum operators related to the three metric $h_{mn}$,
\be
\left[{\sf p}^{kl}(x),\, h_{mn}(y)\right] =
i \delta^k_l\delta^l_n\delta^3(x-y),
\ee
$R$ is the three dimensional curvature scalar, and $\L$ the
cosmological constant. We assume also that the three-space is a compact
smooth manifold without boundaries.

Equation (\ref{diff}) can be easily solved by assuming that the wavefunction
is a function of integrals over space of scalar densities (local
functionals). We will not use here any minisuperspace approximation so 
we take the lowest order local expressions that are diffeomorphism 
invariant. In what follows we will assume that \be
\Psi = \Psi(\cv,\car),
\ee
where
\be
\cv = \int \sqrt{h(x)}d^3x, \;\;\;\;\; \car = \int \sqrt{h(x)} R(x) d^3x.
\ee

However, as it stands, equation (\ref{ham}) is meaningless. First the
functional derivatives act at the same point, so acting on any 
local functional, the first term in (\ref{ham}) produces $\d(0)$. This
problem is resolved by regularization -- in this paper we
use the heat kernel to regularize divergent expressions. At this point one
important remark is in order. The remarkable property of the metric
representation is that the metric operator acts as multiplication. This
means that we do not need to introduce any background metric for
regularization 
purposes (and later work hard to make the final results
background independent.) What we have for granted are eigenvalues of the
metric operator on any state. Strictly speaking, in the heat kernel we
should use metric operators, but these can be identified with their
eigenvalues.

The second problem is related to the choice of ordering of the operator.
The use of additional freedom given by the ordering parameter
$\a$ will be essential in our procedure.

Thus we rewrite eq. (\ref{ham}) as
$$
\left(-\alpha\int d^3y\,  {\cal G}_{klmn}(x) 
K(x,y;t)\frac{\delta}{\delta h_{kl}(x)}
\frac{\delta}{\delta h_{mn}(y)}-\right. $$
\be
\left.(1-\alpha)
\int d^3y\,  \frac{\delta \left({\cal G}_{klmn}(x) K(x,y;t)\right)}
{\delta h_{kl}(y)}
\frac{\delta}{\delta h_{mn}(x)}
+ \mu^2 \sqrt h (2 \Lambda +  R )\right)\Psi =0
, \label{WDW}
\ee
where $K(x,y;t)$ is a heat kernel satisfying the equation
\be
\frac{\partial}{\partial t}K(x,y;t) = \nabla^2_{(x)} K(x,y;t) + \xi
R(x) K(x,y;t) \label{hke}
\ee
with the initial condition
\be
\lim_{t \rar 0} K(x,y;t) = \frac{\delta^{(3)}(x-y)}{\sqrt{h(x)}}.
\ee
Let us note that in (\ref{WDW}) we cannot take a more general ordering 
involving second
functional derivatives of $K$, since these will again produce $\d^{(3)}(0)$.
Equation (\ref{hke}) can be solved pertubatively in powers of $t$. To the order
which will be of interest in the present context, the solution reads \cite{HK}
\ba
&&K(x,y;t)  = \frac{\exp\left(\left(-\frac1{4t}h_{mn}(x) -
\frac1{24}R_{mn}(x)\right)\D^m\D^n\right)}{(4\pi t)^{\frac32}}*\nonumber\\
&&\left( 1 + t\left(\xi - \frac16\right) R(x) +
t^2 \left({\frac 16}\left(\xi -\frac15\right) \Box R(x)+ 
\right.\right.\\
&&\left.\left. {\frac 12}\left(\xi -\frac16\right)^2 R^2(x)+
\frac1{60}R_{mn}(x)R^{mn}(x) -\frac1{180} R^2(x)\right) + O(t^3)\right)
\nonumber
\label{kxyt}
\ea
where $\Delta^m = x^m - y^m$. 
In order to satisfy the Wheeler-De Witt equation we have to cancel 
$\Box R$ in the above expression so we choose $\xi = \frac15$.

We will use below a shorthand notation
\be
K_{klmn}(x,y;t):={\cal G}_{klmn}(x) K(x,y;t)
\ee

We need one more technical result, namely
the value of
\be
\int d^3y\, \frac{\delta K_{klmn}(x,y;t)}{\delta
h_{kl}(y)}.
\label{delkg}
\ee
This value cannot be derived by taking derivatives of terms in
the expansion for $K$ above (it
involves terms of all orders in $t$), but can be found by solving
perturbatively the equation for $\frac{\delta K}{\delta h_{mn}}$. After
straightforward but tedious calculations one finds (to the leading order
in $t$) that the expression (\ref{delkg}) equals
$$
\frac{1}{(4\pi)^{\frac32}} \left(t^{-\frac32}b_1 h_{mn}(x) + 
t^{-\frac12} (b_2 R_{mn}(x) + b_3 h_{mn}(x)R(x)) + O(t^{\frac12})\right).
$$
where
\be
b_1=\frac34 +\frac{\xi}3,\ \ \ b_2=\frac{13\xi}{6},\ \ \ \
b_3=\frac{\xi^2}3-\frac{5\xi}{12}-\frac1{8}
\label{bis}
\ee

Now we are ready to present our main result. 
We assume that 
$\Psi$ depends only on ${\cal V}$ and ${\cal R}$.
Substituting this form 
to the Wheeler -- De Witt equation (\ref{WDW}) we find terms
proportional to
\be
\int d^3y\,  K_{klmn}(x,y;t)\frac{\delta^2 {\cal V}}{%
\delta h_{kl}(x)\delta h_{mn}(y)} =-\frac{21}8 \sqrt{h(x)}K(x,x;t)
\ee
\be
\int d^3y\,  K_{klmn}(x,y;t)\frac{\delta {\cal V}}{\delta 
h_{kl}(x)} \frac{\delta{\cal V}}{\delta h_{mn}(y)} =-\frac38 \sqrt{h(x)}
\ee
\ba
&&\int d^3y\,  \frac{\delta K_{klmn}(x,y;t)}{\delta
h_{kl}(y)}\frac{\delta{\cal V}}{\delta h_{mn}(x)} =\nonumber\\
&&\frac{1}{(4\pi)^{\frac32}} \left(t^{-\frac32} \frac32 b_1 +
t^{-\frac12} (b_2  + 3 b_3 )R(x) + O(t^{\frac12})\right).
\ea
\ba
&&\int d^3y\,  K_{klmn}(x,y;t)\frac{\delta^2 {\cal R}}{%
\delta h_{kl}(x)\delta h_{mn}(y)} =\nonumber\\
&&\sqrt{h(x)}\left(-3\frac{\pa K(x,x;t)}{\pa t}+(3\xi 
+7/8)R(x) K(x,x;t)\right)
\ea
\be
\int d^3y\,  K_{klmn}(x,y;t)\frac{\delta {\cal R}}{\delta 
h_{kl}(x)} \frac{\delta{\cal R}}{\delta h_{mn}(y)} =
\sqrt{h(x)}\left(R_{mn}(x)R^{mn}(x)-\frac{3}8 R^2(x)\right) 
\ee
\be
\int d^3y\,  K_{klmn}(x,y;t)\frac{\delta {\cal R}}{\delta 
h_{kl}(x)} \frac{\delta{\cal V}}{\delta h_{mn}(y)} =
\sqrt{h(x)}\left(-\frac{1}8 R(x)\right)
\ee
\ba
&&\int d^3y\,  \frac{\delta K_{klmn}(x,y;t)}{\delta
h_{kl}(y)}\frac{\delta{\cal R}}{\delta h_{mn}(x)}=\frac{1}{(4\pi)^{\frac32}}
 \left(t^{-\frac32} \frac12 b_1 R(x)+\right.\nonumber\\
&&\left. t^{-\frac12} \left((b_2  + b_3) R^2(x) -b_2 
R_{mn}(x)R^{mn}(x)\right)+ O(t^{\frac12})\right). \ea
where $b_i$ are given by (\ref{bis}).

Now we face the problem as to if renormalize the equation (i.e., replace
the singular terms involving inverse powers of $t$) or to find a solution
of the equation and renormalize the so obtained wave function. Let us
make this first choice. We use the method proposed by Mansfield
\cite{man} which, after multiplying by arbitrary function of $t$ and
analytical continuation results effectively in replacing the
powers
$t^{-\frac p2}$ by $p$th derivatives of arbitrary function $\phi(s)$ at 
$s=0$,
($s=\sqrt{t}$) which we denote by $\phi^{(p)}$. These numbers are therefore 
renormalization constants.

Collecting all terms we have four equations
(the coefficients multiplying $R^2(x)$, $R_{mn}R^{mn}$,
$R(x)$ and $1$ respectively):
\be
\frac38\frac{\pa^2 \Psi}{\pa {\cal R}^2}+
\frac{\pa \Psi}{\pa 
{\cal R}}\po\left( \frac32\a b_3+\frac12\a
b_2-\frac{19}{720} \right)=0
\label{r2}
\ee
\be
-\frac{\pa^2 \Psi}{\pa {\cal R}^2}+
\frac{\pa \Psi}{\pa {\cal R}}\po
\left(-\a b_2 -\frac3{80} \right)=0 
\label{rmn2}
\ee
\be
\frac14\frac{\pa^2 \Psi}{\pa {\cal R}\pa \cv}+
\frac{\pa \Psi}{\pa 
{\cal R}}\pt\left( \frac12\a b_1 -\frac58\right)
+\frac{\pa \Psi}{\pa \cv}\po\left(\frac{7}{80}
+\frac32\a b_3 +\frac12\a b_2\right)+\m^2\Psi=0
\label{lr}
\ee
\be
\frac38\frac{\pa^2 \Psi}{\pa {\cv}^2}+
\frac{\pa \Psi}{\pa 
{\cv}}\pt\left( \frac{21}8+\frac32\a
b_1 \right)
+\frac{\pa \Psi}{\pa 
{\car}}\frac{3\pf}4
+2\L\m^2\Psi=0
\label{l1}
\ee

These equations are linear differential equations with
constant coefficients so the solutions are
combinations of exponential functions of $\cv$ and $\car$.

We will present an explicit solution 
when the wavefunction depends only on $\cv$ (the general
case is not difficult but the equations are rather lengthy).
In this case 
only two of the equations survive and we substitute
\be
\Psi({\cal V}) = Ae^{\omega_1 {\cal V}},
\ee
Then we get from (\ref{lr}) and (\ref{l1})
that $\omega_1$ satisfies
\be
\frac38\omega_1^2+\frac{21}8
\phi^{(3)}\omega_1 +2\L\m^2-\frac{3 b_1 \phi^{(3)}}{\phi^{(1)}(3b_3+b_2)}
\left(\frac7{80}\phi^{(1)}\o_1+\m^2\right)=0
\label{om1}
\ee
The dependence of the solution on $\mu^2\sim G^{-2}$ can be written as
\be
\omega_1=-B\pm\sqrt{B^2-C\mu^2}
\label{omjed}
\ee
where the constants $B$ and $C$ can be extracted from eq. (\ref{om1}).
Depending on the values of the constants we
get real or complex wave functions - the interpretation in these cases 
would be quite different.

In the most general case when the function depends on both 
$\cv$ and $\car$ we get a relation between the cosmological constant and 
the Newton's constant expressed via the renormalization constants.

\vskip.5cm
We conclude this paper with a number of comments:
\begin{enumerate}
\item
It is natural to ask if one can find a solution depending on
higher order local functionals like e.g. $S_2=\int (
bR^2+cR_{mn}R^{mn})$. The
procedure should be as follows. First, in order to avoid
terms of order $t^3$ in the heat kernel expansion, we must
choose $b$ and $c$ such as to avoid fourth
order covariant derivatives in the second functional
derivative of $S_2$ (otherwise we would have a large number
of independent third order tensor structures rather
impossible to match). This gives one relation between
the coefficients. Then one should try to match terms with
independent combinations of curvature and metric tensors up
to the order three. It is unclear at the moment if this
procedure leads to any nontrivial solution.
\item
There are several directions of obvious generalizations of
our result that are currently under investigations. First,
one may consider inclusion of matter fields. The case of a
scalar field and the closely related analysis of the
cosmological constant problem will be the subject of the
forthcomming paper. Second, it is of outmost importance for
e.g., black holes physics to consider a theory based on a
compact manifold with boundaries. This problem is, however,
much more complicated.
\item
It should be stressed that the wave function alone does not
provide us with much insight into the physics of the
solution. We certainly lack one important ingredient which
is an inner product. In context of the solutions presented
above, our goal will be to write down the path integral
measure $\mu(h_{mn})dh_{mn}$ in the form
$$
dh_{mn}^{\bot}d{\cal V}d{\cal R}\ J(h^{\bot},{\cal V},{\cal R}),
$$
where $\delta h^{\bot}_{mn}$ are variations of the metric
which leave ${\cal V}$ and ${\cal R}$ invariant:
$$
h^{mn}\d h^{\bot}_{mn}=0,\ \ R^{mn}\d h^{\bot}_{mn}=0
$$
and then integrate over $ h^{\bot}_{mn}$ to obtain
$$
\m(\cv,\car)d\cv d\car.
$$
Only after succeeding this program we will be able to
address questions concerning hermiticity of operators and
their expectation values.

Alternatively, we may make use of the quantum potential
approach to quantum gravity which provides us with the four
dimensional dynamics related to a given wave function
\cite{J}. The major virtue of this approach is that most
physical predictions concerning the evolution do not depend
on the inner product.
\item
Given the regularized WDW operator we can analyse the issue
of anomalies. This problem has been investigated in
\cite{3jap}, but we found ourselves in disagreement with the
results of the papers.
There are also additional, potentially anomalous terms
resulting from the general ordering chosen in our approach.
We are currently investigating this problem, and the result
will be published soon.
\item
One may wonder why such simple solutions were never found in
spite of many papers on semiclassical expansion of the WDW
equation. The reason is that in the semiclassical expansion,
as a rule, one employs an expansion of the logarithm of the 
wave function in powers of $G^{-2}$. 
However, as our result shows, the exact
solution is of the form e.g. $e^{\omega_1\cv}$ with $\omega_1$ being
a nonpolynomial function of $G^{-2}$ (eq. (\ref{omjed}))
therefore such an expansion is not finite (one should recall
that the renormalization coefficients $\phi^{(n)}$ are
dimensionful and therefore can mix with powers of $G^{-2}$).
\end{enumerate}

\end{document}